\documentstyle[12pt,psfig]{article}
\def\ii{\'{\char'20}}

\begin{document}

\newcommand{\ep}{\epsilon}
\newcommand{\fr}{\frac}
\newcommand{\reals}{\mbox{${\rm I\!\!R }$}}
\newcommand{\nats}{\mbox{${\rm I\!\!N }$}}
\newcommand{\intgs}{\mbox{${\rm Z\!\!Z }$}}
\newcommand{\cam}{{\cal M}}
\newcommand{\caz}{{\cal Z}}
\newcommand{\cao}{{\cal O}}
\newcommand{\cac}{{\cal C}}
\newcommand{\aaa}{\int\limits_{mR}^{\infty}dk\,\,}
\newcommand{\bbb}{\left[\left(\frac k R\right)^2-m^2\right]^{-s}}
\newcommand{\ccc}{\frac{\partial}{\partial k}}
\newcommand{\fff}{\frac{\partial}{\partial z}}
\newcommand{\iikma}{\aaa \bbb \ccc}
\newcommand{\abc}{\int\limits_m^{\infty}dk\,\,}
\newcommand{\eef}{[k^2-m^2]^{-s}}
\newcommand{\imk}{\abc\eef\ccc}
\newcommand{\ddd}{\int\limits_{mR/\nu}^{\infty}dz\,\,}
\newcommand{\eee}{\left[\left(\frac{z\nu} R\right)^2-m^2\right]^{-s}}
\newcommand{\lll}{\frac{(-1)^j}{j!}}
\newcommand{\iinma}{\ddd\eee\fff}
\newcommand{\cah}{{\cal H}}
\newcommand{\nn}{\nonumber}
\renewcommand{\theequation}{\mbox{\arabic{section}.\arabic{equation}}}
\newcommand{\komplex}{\mbox{${\rm I\!\!\!C }$}}
\newcommand{\sip}{\frac{\sin (\pi s)}{\pi}}
\newcommand{\numr}{\left(\frac{\nu}{mR}\right)^2}
\newcommand{\mzs}{m^{-2s}}
\newcommand{\rzs}{R^{2s}}
\newcommand{\abl}{\partial}
\newcommand{\ablz}{{Z_D^{\nu}}'(0)}
\newcommand{\g}{\Gamma\left(}
\newcommand{\zzz}{\int\limits_{\gamma}\frac{dk}{2\pi i}\,\,}
\newcommand{\yyy}{(k^2+m^2)^{-s}\frac{\partial}{\partial k}}
\newcommand{\ikma}{\zzz \yyy}
\newcommand{\ead}{e_{\alpha}(D)}
\newcommand{\sual}{\sum_{\alpha =1}^{D-2}}
\newcommand{\sulnu}{\sum_{l=0}^{\infty}}
\newcommand{\sujnu}{\sum_{j=0}^{\infty}}
\newcommand{\suani}{\sum_{a=0}^i}
\newcommand{\suanzi}{\sum_{a=0}^{2i}}
\newcommand{\zend}{\zeta_D^{\nu}}
\newcommand{\amed}{A_{-1}^{\nu ,D}(s)}
\renewcommand{\and}{A_{0}^{\nu ,D}(s)}
\newcommand{\aid}{A_{i}^{\nu ,D}(s)}
\def\beq{\begin{eqnarray}}
\def\eeq{\end{eqnarray}}

\begin{titlepage}

\title{\begin{flushright}
{\normalsize UB-ECM-PF 95/10, hep-th/9505157}
\end{flushright}
\vspace{3mm}
{\Large \bf Zeta function determinant of the Laplace operator on the
$D$-dimensional ball}}
\author{M. Bordag\thanks{E-mail address:
bordag@qft.physik.uni-leipzig.d400.de},
B. Geyer\thanks{E-mail address:
geyer@ntz.uni-leipzig.d400.de},
K. Kirsten\thanks{E-mail address: kirsten@tph100.physik.uni-leipzig.de}\\
Universit{\"a}t Leipzig, Institut f{\"u}r Theoretische Physik,\\
Augustusplatz 10, 04109 Leipzig, Germany\\
\\
E. Elizalde\thanks{E-mail address: eli@zeta.ecm.ub.es}\\
CEAB, CSIC, Cam\'{\i} de Santa
B\`arbara,
17300 Blanes,
\\ and Departament d'ECM and IFAE, Facultat de F{\ii}sica,\\
Universitat de Barcelona, Av. Diagonal 647, 08028 Barcelona\\
Spain}

\thispagestyle{empty}

\vspace*{-1mm}

\maketitle

\vspace*{-2mm}

\begin{abstract}
We present a direct approach for the calculation
of functional determinants of the Laplace operator on balls.
Dirichlet and Robin boundary conditions are considered. Using
this approach, formulas for any value of the dimension, $D$,
of the ball, can be obtained quite easily.
Explicit results are presented here for dimensions
$D=2,3,4,5$ and $6$.
\end{abstract}
\end{titlepage}
\section{Introduction}
\setcounter{equation}{0}
Motivated by the need to give answers to some fundamental questions in
quantum field theory, during the
 last years there has been (and continues to be) a lot of interest in the
problem of calculating the determinant of a differential operator, $L$
(see
for example \cite{ramond81,birelldavies82}). Often one has to deal
in these situations with
positive elliptic differential operators acting on sections of a vector
 bundle
over a compact manifold. In such cases $L$ has a discrete spectrum
$\lambda_1\leq
\lambda_2\leq ...\to\infty$. The determinant, $\det L=\prod_i\lambda_i$,
 is
generally divergent and one needs to make sense out of it by
means of some kind of analytic continuation.
A most appropriate way of doing that is by using  the zeta
function regularization prescription introduced by Ray and Singer
\cite{raysinger71}
(see also \cite{critchleydowker76,hawking75}). In this procedure
 $ \ln \det L$ is defined
by analytically continuing the function $\sum_i \lambda_i^{-s} \ln\lambda_i$
in the exponent $s$, from the domain of the complex plane where
the real part of $s$ is large to the point $s=0$.
Introducing the zeta function associated with the spectrum $\lambda_i$
of $L$,
\beq
\zeta (s) =\sum_i \lambda_i^{-s},\nn
\eeq
this is equivalent to defining
\beq
\ln\det L =-\zeta ' (0).\nn
\eeq
Only a few general methods for the exact evaluation of $\ln\det L$ are
available. Thus, for example, given that the
manifold has a boundary, in \cite{forman92} (see also
 \cite{dreyfussdym78,forman87,forman89,burgheleafriedlanderkappeler92})
the determinants of differential and difference operators have been
 related
to the boundary values of solutions of the operators.
When $L$ is a conformally covariant differential operator, exact
results may sometimes
 be obtained by transforming to a ``more simple" operator $\tilde L$ for
which $\ln\det \tilde L$ is known. Then, the knowledge of the associated
heat-kernel coefficients ---nowadays available
 \cite{bransongilkey90,gilkeynew}---
gives sometimes  the exact value
of $\ln\det L$
\cite{dowkerschofield90,blauvisserwipf89,blauvisserwipf88a}.
This approach has been used by Dowker to find the functional determinants for
a variety of sectors of Euclidean space, spheres and flat balls for dimensions
$D\leq 4$ \cite{dowker94a,dowker94b,dowker94c}.
Similar techniques
have proven to be very powerful in order to obtain estimates of different
types
\cite{osgoodphillipssarnak88,osgoodphillipssarnak88a,bransonchangyang92}.

As a rule, however, explicit knowledge of the eigenvalues $\lambda_i$ is
necessary in order to obtain exact results for $\ln\det L$. This explicit
knowledge of the eigenvalues is in general only guaranteed
 for highly symmetric regions of space,
such as the torus, sphere or regions bounded by parallel planes. For these
manifolds, detailed calculations have been performed in the context of
Casimir energies and effective potential considerations (for a summary of
results see \cite{eorbz}).

In this paper we want to focus on a class of situations
 for which the eigenvalues of the
operator are not known explicitly, but nevertheless the exact calculation of
$\ln\det L$ is possible. The method developed is applicable whenever an
implicit
equation satisfied by the eigenvalues is known and some properties
 (later specified) of this equation are known.
 We exemplify our approach by taking $L=-\Delta$ on
the $D$-dimensional ball $B^D =\{x\in \reals ^D; |x|\leq R\}$, together with
Dirichlet ---or general Robin--- boundary conditions.
In dimensions $D\leq 5$ a
part of these results may also be obtained using the conformal
transformation techniques mentioned above. (Here the restricition $D\leq
5$ results from the
need of the knowledge of the heat-kernel coefficients $a_{D/2}$, which has
been determined only recently for $D=5$ \cite{gilkeynew}.) This particular
method was
used by Dowker and Apps for $D\leq 4$ (this was before $a_{5/2}$ were
known) \cite{fundow}. Here
we will apply a direct approach which has already been shown to be quite
 powerful
for the calculation of the heat-kernel coefficients in the situations
described above \cite{bek}.
Interesting quantum field theoretical applications
of the results obtained can be found in
quantum cosmology
\cite{barvinskykamenshchikkarmazin92}
\cite{kamenshchikmishakow92,deathesposito91,louko88,schleich85}.
For these applications,  the consideration of dimensions $D$ higher than
four
are of interest, because the technicalities involved in dealing with
higher-spin fields
reduce essentially to the ones for scalar fields in those dimensions.
Further applications are in statistical mechanics, in connection with
finite size effects
\cite{capellicosta89}, and in conformal field theory
\cite{guraswamyrajeevvitale}.

The paper is organized as follows.
In section 2 we describe in detail our approach for the calculation of
functional determinants using as an example the Laplace operator of the
three-dimensional ball with Dirichlet boundary conditions. For
generality and
because the analytic continuation procedure employed is slightly easier, we
start with the massive Laplacian, performing the limit $m\to 0$ at a suitable
point of our calculation. The result we find here agrees completely with
the one recently given
by Dowker and Apps \cite{fundow}. After having explained in detail the
main ideas, we
apply our approach to an (in principle) arbitrary dimension $D$ and also
to
general Robin boundary conditions. We start in Sect. 3 with Robin
boundary
conditions for the three-dimensional ball. The Neumann boundary
conditions ---a
special case of the Robin boundary conditions--- cannot be obtained as a
limit of the parameter involved in Robin boundary conditions. An extra
consideration,
necessary in order to deal with this situation, is given at the end of
section 3. In Sect. 4 we describe in detail how our scheme
can be applied in dimensions $D>3$. Explicit results are
given for dimensions $D=4,5,6$ and
all the different boundary conditions. The case $D=2$ is briefly
considered in Sect. 5. This case
is slightly different because the lowest angular momentum $l=0$ needs to be
treated in a specific way. In the conclusions (Sect. 6) we summarize the
main results of our investigation.
As we will see in the course of our procedure,
 the functional determinant is
naturally splitted up into two pieces. The contributions of each of
these
pieces and a detail of how they are obtained are given in three
appendixes.
\section{Zeta function determinant on the $3$-dimensional ball with
Dirichlet boundary conditions}
\setcounter{equation}{0}
In this chapter we want to concentrate on the $3$-dimensional ball with
Dirichlet boundary conditions in order to exemplify our direct approach,
which is actually
applicable in any dimension and  for completely general Robin boundary
condition. We are thus interested in obtaining the zeta function
of the operator $(-\Delta +m^2)$ on the ball
$B^3=\{x\in\reals ^3; |x|\leq R\}$ endowed with Dirichlet
boundary conditions. The zeta function is formally defined as
\beq
\zeta (s) =\sum_k \lambda_k ^{-s},\label{eq:2.1}
\eeq
with the eigenvalues $\lambda_k$ being determined through
\beq
(-\Delta +m^2) \phi_k (x) =\lambda_k \phi _k (x)\label{eq:2.2}
\eeq
($k$ is in general a multi-index here), together with the
boundary
condition.
It is convenient to introduce a spherical coordinate basis, with $r=|x|$
and
angles $\Omega =(\theta,\varphi )$.
In these coordinates, a complete set of solutions of Eq.~(\ref{eq:2.2})
can be given in
the form
\beq
\phi_{l,m,n}(r,\Omega )=r^{-\frac 1 2} J_{l+\frac 1 2}
(w_{l,n}r) Y_{l+\frac 3 2} (\Omega ),\label{eq:2.3}
\eeq
the $J_{l+1/2}$ being Bessel functions and
the $Y_{l+3/2}$ hyperspherical harmonics
\cite{erdelyimagnusoberhettingertricomi53}. The $w_{l,n}$
$(>0)$ are determined through the boundary condition by
\beq
J_{l+\frac{1} 2 }(w_{l,n}R) &=& 0.\label{eq:2.4}
\eeq
In this notations, using $\lambda_{l,n}=w_{l,n}^2+m^2$, the zeta
function can be given in the form
\beq
\zeta (s) =\sum_{n=0}^{\infty} \sulnu (2l+1) (w_{l,n}^2
+m^2)^{-s}.\label{eq:2.5}
\eeq
Here the sum over $n$ is extended over all possible solutions of
Eq.~(\ref{eq:2.4}) on the positive real axis, and $(2l+1)$ is the number
of independent
harmonic polynomials
in $3$ dimensions.

As it stands, the zeta function in Eq.~(\ref{eq:2.5}) is defined for
$\Re s > 3/2$. The way how to construct the analytical continuation of
Eq.~(\ref{eq:2.5}) to the left has been explained in detail in
Ref. \cite{bek} for the calculation of the heat-kernel coefficients
associated with the operator (\ref{eq:2.2}). For this reason our
description here will be brief. Starting point of the consideration is
the representation of $\zeta (s)$ by the contour integral
\beq
\zeta (s) =\sulnu (2l+1) \ikma\ln J_{l+\frac 1 2} (kR).\label{eq:2.6}
\eeq
where the contour $\gamma$ runs counterclockwise and must enclose all
solutions of Eq.~(\ref{eq:2.4}) on the positive real axis (for a similar
treatment as a contour integral see
\cite{kamenshchikkarmazin92,bordag95,barvinskykamenshchikkarmazin92}).
Subtracting and adding the leading asymptotic terms of $I_{\nu} (z\nu )$
for $\nu \to \infty$, $\nu =l+1/2$, the following representation of
$\zeta (s)$ valid in the strip $-1/2 <\Re s <1$ is found to be valid (for
details see \cite{bek}),
\beq
\zeta (s) =2 \sulnu \nu Z_D^{\nu}(s) +\sum_{i=-1}^{2}A_i^
D(s),\label{eq:2.7} \eeq
with the definitions
\beq
Z_D^{\nu}(s) &=& \sip \left. \iinma\right\{\ln\left[I_{\nu} (z\nu
)\right]\label{eq:2.7a}\\
& &\hspace{3cm}\left.
-\ln\left[\frac{1}{\sqrt{2\pi\nu}}\frac{e^{\nu\eta}}
{(1+z^2)^{\frac
1 4}}\right]-\sum_{n=1}^2 \frac{D_n (t)}{\nu ^n}\right\},\nn
\eeq
\beq
A_{-1}^D (s) =\frac{\rzs}{2\sqrt{\pi}\Gamma (s)}\sujnu \lll
(mR)^{2j}\frac{\g j+s-\frac 1 2\right)}{s+j}\zeta_H (2j+2s
-2;1/2),\label{eq:2.7b}
\eeq
\beq
A_0^D (s) =-\frac{\rzs}{2\Gamma (s)}\sujnu \lll (mR)^{2j}\Gamma (s+j)
\zeta_H (2j+2s-1;1/2),\label{eq:2.7c}
\eeq
and, for $i=1,2$,
\beq
A_i^D (s) &=& -\frac{\rzs}{\Gamma (s)}\sujnu \lll (mR)^{2j} \zeta_H
(-1+i+2j+2s;1/2) \nn\\
& &\hspace{3cm} \times \suani x_{i,a} \frac{(i+2a)\g s+a+j+\frac i
2\right)}{\g 1+a+\frac i 2\right)}.\label{eq:2.7d}
\eeq
Here
\beq
D_1 (t) &=& \sum_{a=0}^1 x_{1,a} t^{1+2a}=\frac 1 8 t-\frac 5 {24}
t^3,\nn\\
D_2 (t) &=& \sum_{a=0}^2 x_{2,a} t^{1+2a}=\frac 1 {16} t^2-\frac  {8}
t^4+\frac 5 {16} t^6,\nn
\eeq
and $t=1/\sqrt{1+z^2}$, $\eta =\sqrt{1+z^2}+\ln (z/1+\sqrt{1+z^2})$.
Eq.~(\ref{eq:2.7}) is a very suitable starting point for the calculation
of the zeta function determinant $\zeta '(0)$; here we consider, for
definiteness, $m=0$. In the limit $m\to 0$ only the $j=0$ term survives
in the terms $A_n^D (s)$, $n=-1,0,1,2$, and one immediately finds
\beq
\left. \frac d {ds}\sum_{i=-1}^2A_i^D(s)\right|_{s=0}=-\frac 3 {32}
+\frac{\ln 2}{24} -\frac{\ln R}{24}+\frac 3 2 {\zeta_R}'(-2) +\frac 1 2
{\zeta_R}' (-1).\label{eq:2.8}
\eeq
For the part $Z_D^{\nu}(s)$ some additional calculation is needed. First of
all using the analyticity of $Z_D^{\nu} (s) $  around $s=0$, the derivative
${Z_D^{\nu}}'(0)$ is  found to be
\beq
\left.
\ablz =-\left[ \ln I_{\nu} (mz) -\nu\eta+\ln\left(\sqrt{2\pi
\nu}(1+z^2)^{1/4}\right)-\frac{D_1(t)}{\nu}
-\frac{D_2(t)}{\nu^2}\right]\right|_{z=(mR)/\nu},\label{eq:2.9}
\eeq
and, in the limit $m\to 0$, this reduces to
\beq
\ablz =\ln \Gamma (\nu +1) +\nu -\nu \ln \nu -\frac 1 2 \ln (2\pi \nu
)-\frac 1 {12\nu}.\label{eq:2.10}
\eeq
To perform afterwards the sum over $\nu$, it is very convenient to use
the integral representation of $\ln\Gamma (\nu +1)$
\cite{gradshteynryzhik65}, to find
\beq
\ablz =\int\limits_0^{\infty}dt\,\, \left(-\frac t {12} +\frac 1 2
-\frac 1 t +\frac 1 {e^t-1}\right)\frac{e^{-t\nu}} t.\label{eq:2.11}
\eeq
Performing the sum, this yields
\beq
Z_D'(0) =2\int\limits_0^{\infty}dt\,\, \left(-\frac 1
{2t^2}+\frac 2 {t^3} +\frac d {dt} \frac{e^{-t}}
{t(1-e^{-t})}\right)\frac{e^{-\frac t 2}}{1-e^{-t}}.\label{eq:2.12}
\eeq
As is seen by studying the asymptotics for $t\to 0$, the integral is
well defined. For the explicit calculation of $Z_D'(0)$,
Eq.~(\ref{eq:2.12}), it is suitable to introduce a regularisation
parameter and to define
\beq
Z_D'(0,z) =2\int\limits_0^{\infty}dt\,\, t^z\left(-\frac 1
{2t^2}+\frac 2 {t^3} +\frac d {dt} \frac{e^{-t}}
{t(1-e^{-t})}\right)\frac{e^{-\frac t 2}}{1-e^{-t}},\label{eq:2.13}
\eeq
with
\beq
Z_D'(0,0) =Z_D'(0).
\eeq
The individual pieces of the integral Eq.~(\ref{eq:2.13}) may then be
calculated by means of \cite{gradshteynryzhik65}
\beq
\int\limits_0^{\infty}dx\,\,\frac{x^{\nu-1}e^{-\mu x}}{1-e^{-\beta
x}}=\frac{\Gamma (\nu )}{\beta^{\nu}}\,\zeta_H \left(\nu;\frac{\mu}{
\beta}\right),\label{eq:2.14}
\eeq
and differentiating with respect to $\beta$,
\beq
\int\limits_0^{\infty}dx\,\,\frac{x^{\nu}e^{-(\mu+\beta)
x}}{(1-e^{-\beta x})^2}=\frac{\Gamma (\nu +1 )}{\beta^{\nu +1}}\,\zeta_H
\left(\nu;\frac{\mu}{ \beta}\right)
-\frac{\Gamma (\nu +1 )\mu}{\beta^{\nu +2}}\,\zeta_H
\left(\nu +1;\frac{\mu}{ \beta}\right).
\label{eq:2.15} \eeq
As a result we have
\beq
Z_D'(0,z) =\zeta_H \left(z-2;\frac 1 2\right)\Gamma (z-2) [6+z-z^2]
+\frac 1 4 \zeta_H \left(z;\frac 1 2 \right)\Gamma (z)\nn
\eeq
and thus
\beq
Z_D'(0) =-\frac 9 4 \zeta_R'(-2) -\frac 1 8 \ln 2. \label{eq:2.16}
\eeq
Adding up the contributions from the Eq.~(\ref{eq:2.8}) and
(\ref{eq:2.16}) we end up with
\beq
\zeta'(0)=-\frac 3 {32} -\frac 1 {12}\ln 2 -\frac 3 4
\zeta_R'(-2)+\frac 1 2 \zeta_R'(-1) -\frac 1 {24} \ln R\label{eq:2.17}
\eeq
in agreement with Dowker and Apps \cite{fundow}.
\section{Zeta function determinant on the $3$-dimensional ball with
Robin boundary conditions} \setcounter{equation}{0}
In order to treat Robin boundary conditions, only very little changes
are necessary. Writing the boundary condition in the form
\beq
\frac u R J_{l+1/2} (\omega_{l,n}R) +w_{l,n}J_{l+1/2}'(w_{l,n}r)
\left|_{r=R} =0,\right.\label{eq:3.1}
\eeq
the starting point of the calculation, analogous to
Eqs.~(\ref{eq:2.7a})-(\ref{eq:2.7d}), is
\beq
Z_R^{\nu}(s) &=& \sip \iinma\left\{\ln\left[\frac u R I_{\nu} (z\nu
)+\frac{z\nu}{R}I_{\nu}'(\nu z)\right]\right.\label{eq:3.2}\\
& &\hspace{3cm}\left.
-\ln\left[\frac{\nu}{R\sqrt{2\pi\nu}}e^{\nu\eta}
(1+z^2)^{\frac
1 4}\right]-\sum_{n=1}^2 \frac{M_n (t)}{\nu ^n}\right\},\nn
\eeq
\beq
A_{-1}^R (s) =A_{-1}^D(s),\label{eq:3.3}
\eeq
\beq
A_0^R (s) =- A_0^D(s),\label{eq:3.4}
\eeq
and for $i=1,2$,
\beq
A_i^R (s) &=& -\frac{\rzs}{\Gamma (s)}\sujnu \lll (mR)^{2j} \zeta_H
(-1+i+2j+2s;1/2) \nn\\
& &\hspace{3cm} \times \sum_{a=0}^{2i} z_{i,a} \frac{(i+a)\g
s+j+\frac{a+ i} 2\right)}{\g 1+\frac {a+i} 2\right)}.\label{eq:3.5}
\eeq
Here we need the polynomials
\beq
M_1 (t,u) &=& \sum_{a=0}^2 z_{1,a} t^{1+a}=\left(-\frac 3 8
+u\right)t+\frac 7 {24} t^3,\label{eq:3.6}
\eeq
and
\beq
M_2 (t,u) &=& \sum_{a=0}^4 z_{2,a} t^{2+a}=\left(-\frac 3 {16}+\frac u 2
-\frac{u^2} 2\right) t^2+\left(\frac 5 8 -\frac u 2\right) t^4-\frac 7
{16} t^6.\label{eq:3.7} \eeq
The contribution from the asymptotic terms is
\beq
\left. \frac d {ds}\sum_{i=-1}^2A_i^R(s)\right|_{s=0}
&=&\frac 1 {32}
-\frac{\ln 2}{24} +\frac{\ln R}{24}+\frac 3 2 {\zeta_R}'(-2) -\frac 1 2
{\zeta_R}' (-1)\nn\\
& &+\frac u 2 +\gamma u^2 +2u^2\ln 2 +u^2 \ln R.\label{eq:3.8}
\eeq
For the rest, things are very similar to the Dirichlet case.
We find
\beq
{Z^{\nu}_R}'(0) &=&\int\limits_0^{\infty}dt\,\, \left(-\frac t {12}
+\frac 1
2 -\frac 1 t +\frac 1 {e^t-1}\right)\frac{e^{-t\nu}} t\label{eq:3.9}\\
& &+\ln\left(\frac{\nu}{u+\nu}\right)+\frac u {\nu}-\frac{u^2}{2\nu^2},\nn
\eeq
and thus, using Eq.~(\ref{eq:2.11}), we have
\beq
Z_R'(0) = -\frac 9 4 \zeta_R'(-2) -\frac 1 8 \ln 2 +2\sulnu
\left\{-\nu\ln \left(1+\frac u {\nu}\right) +u
-\frac{u^2}{2\nu}\right\}.\label{eq:3.10} \eeq
The remaining sum may be done by differentiating with respect to $u$ and
integrating back. As a result
\beq
\lefteqn{
\sulnu \left\{-\nu\ln \left(1+\frac u {\nu}\right) +u-\frac
{u^2}{2\nu}\right\} }\label{3.11}\\
& &=-\frac 1 2 u^2 (\gamma +2\ln 2) -u\ln\Gamma\left(\frac 1 2
+u\right)+\int\limits_0^u dx\,\,\ln\Gamma\left(\frac 1 2 +x\right).\nn
\eeq
Let us mention that the last equation can also be given in terms of the
Barnes $G$-function \cite{gradshteynryzhik65}.

Putting things together, we arrive at our final result,
\beq
\zeta'(0,u)&=& \frac 1 {32} -\frac 1 6 \ln 2 -\frac 3 4 \zeta_R'(-2)-\frac
1 2 \zeta_R'(-1) +\frac 1 {24} \ln R\label{eq:3.12}\\
& &+\frac u 2 -2u\ln\Gamma\left(\frac 1 2 +u\right)  +u^2 \ln R
+2\int\limits_0 ^u dx\,\,\ln\Gamma\left(\frac 1 2 +x\right)\nn,
\eeq
which concludes the consideration of the three dimensional case with
Robin boundary conditions.
The dependence of $\zeta' (0,u)$ on the parameter $u$ is shown in
Fig. 1 for $R=1$.

\unitlength1cm
\begin{figure}[ht]
\vspace*{-3cm}
\centerline{\psfig{figure=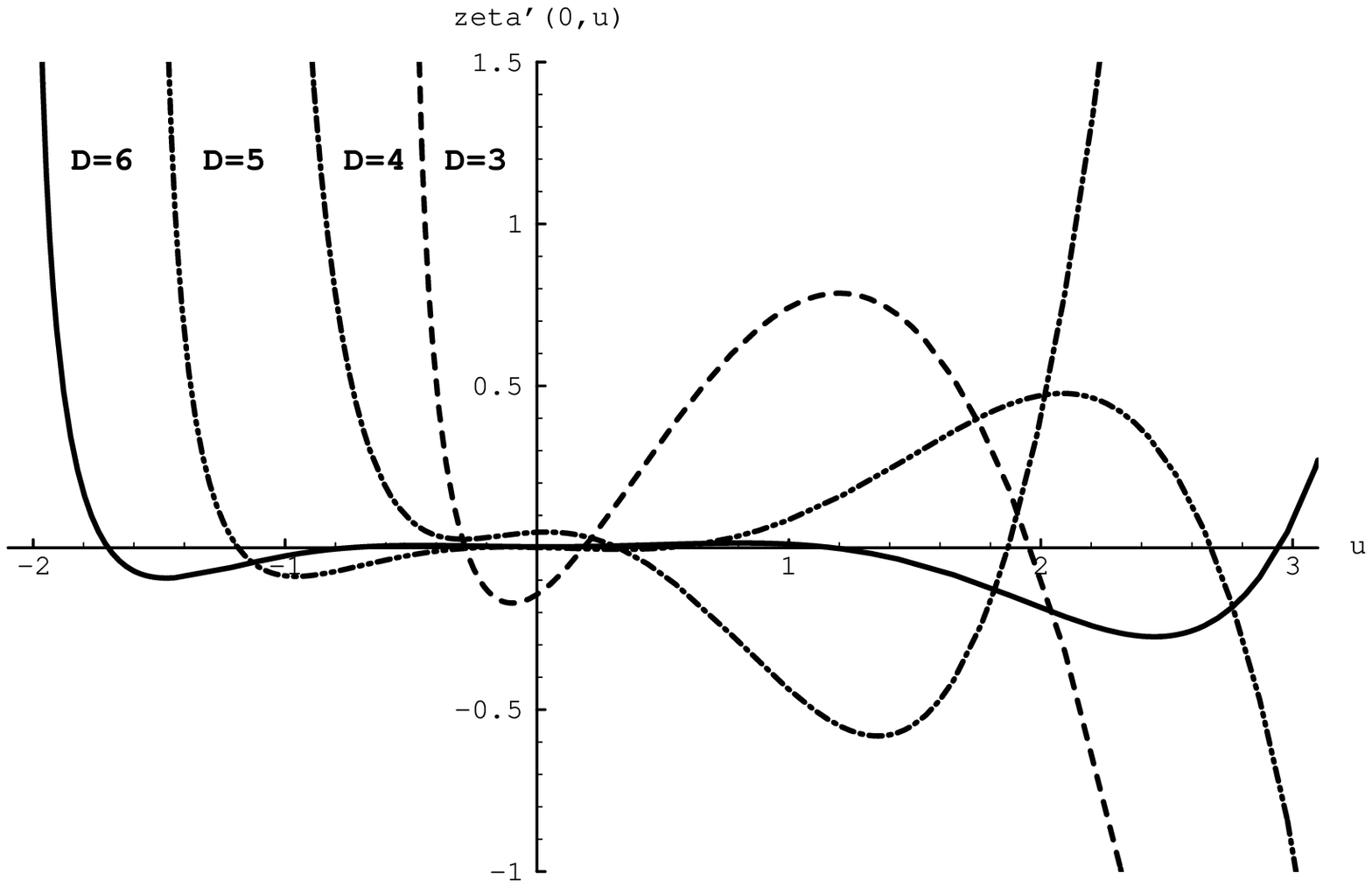,height=14cm,width=12cm}}
\vspace{-3cm}

\noindent{\bf Figure 1:}
\parbox[t]{12cm}{Plot of the dependence of
$\zeta' (0,u)$ on the parameter $u$
 for $R=1$ and for dimensions $D=3,4,5,6$. Notice the divergence that
apppears for $u=1-D/2$ in each dimension, corresponding to the case of
Neumann boundary conditions.}
\vspace{1cm}
\end{figure}

Eq.~(\ref{eq:3.12}) can be given in an
alternative form using the following expression, first obtained in
 \cite{elizalderomeo90b}  (see also \cite{eorbz})
\beq
\int\limits_0^u dx\,\,\ln\Gamma\left(\frac 1 2 +x\right) &=&\zeta_H
(-1,u+1/2) +{\zeta_H}'(-1,u+1/2)\nn\\
& &+\frac u 2 \ln(2\pi )+\frac{\ln 2
-1}{24} +\frac 1 2 {\zeta_R}'(-1).\label{lngam}
\eeq
As is seen, the limit $u\to
-1/2$, corresponding to Neumann boundary conditions, is not smooth,
since
a logarithmic divergence appears due to the $\ln \Gamma (1/2+u)$
term.
This logarithmic divergence might be traced back to the $\nu =1/2$ term
in Eq.~(\ref{eq:3.9}). Thus, in order to find the functional
determinant corresponding to Neumann boundary condition, the term $\nu
=1/2$ has to be treated separately. In fact, looking in detail at the
derivation of Eq.~(\ref{eq:3.9}) starting from Eq.~(\ref{eq:3.2}), it is
seen, that the case $u=-1/2$ has to be treated specifically,
because the behavior of $(u/R)I_{\nu} (z\nu )+(z\nu/R)I_{\nu}'(z\nu)$
for $z\to 0$ is different for $u=-1/2$.

Probably the easiest way to find the results for the Neumann boundary
conditions is to write
\beq
\zeta (s,-1/2)&=&\zeta^{l=0} (s,-1/2)+
 \lim_{u\to -1/2}\left( \zeta (s,u) -\zeta^{l=0} (s,u)\right),\label{neu1}
\eeq
because then we can use all the results for the Robin boundary conditions
that we have derived before. Here $\zeta^{l=0} (s,u)$ is the
contribution from the angular momentum component $l=0$ to $\zeta (s,u)$,
\beq
\zeta ^{l=0}(s,u) &=&\sip \imk\ln\left[\frac u R I_{1/2} (kR)+kI_{1/2}'
(kR)\right].\label{neu2}
\eeq
Proceeding with the calculations as for the Robin boundary conditions, one
easily finds
\beq
\frac d {ds} \left.\left(\zeta^{l=0} (s,-1/2)-\zeta^{l=0} (s,u)\right)
\right|_{s=0}=
 -2\ln R +\ln (3/2) +\ln (u+1/2). \label{neu4}
\eeq
In the limit $u\to -1/2$, the logarithmic divergence in Eq.~(\ref{neu4})
cancels the divergence in $\zeta (s,u)$, Eq.~(\ref{eq:3.12}). Thus the limit
$u\to -1/2$ is well defined and the functional determinant for Neumann boundary
conditions reads
\beq
{\zeta '} (0,-1/2) &=& -{7\over {32}} - {{7\,\ln 2}\over 6} + \ln 3 -
   {{41\,\ln R}\over {24}} - {{3\,{\zeta_R}'(-2)}\over 4} -
   {{{\zeta_R}'(-1)}\over 2}.\nn\\
 & &-2\int\limits_0^{1/2}dx\,\,\ln\Gamma (x). \label{neu3}
\eeq
This concludes the consideration of the $3$-dimensional case: all ordinary
boundary conditions have been dealt with explicitly. As for the Robin
boundary condition, an alternative form may be presented using
\beq
\int\limits_0^{1/2}dx\,\,\ln\Gamma (x) =\frac 1 8 +\frac 5 {24} \ln 2
+\frac 1 4 \ln \pi -\frac 3 2 {\zeta_R}'(-1).\nn
\eeq
\section{Zeta function determinants on $D>3$ dimensional ball}
\setcounter{equation}{0}
For higher dimensional balls, exactly the same procedure may be employed
in order to calculate the zeta function determinant. The starting point is
here (for details see \cite{bek})
\beq
\zeta_D(s) =\sulnu d_l (D) \ikma \ln J_{l+\frac{D-2} 2} (kR),
\label{eq:4.1}
\eeq
with the number $d_l (D)$ of independent harmonic polynomials,
explicitly
\beq
d_l(D) =(2l+D-2) \frac{(l+D-3)!}{l!(D-2)!}.\label{eq:4.2}
\eeq
It is suitable to introduce the coefficients $e_{\alpha}(D)$ by
\beq
d_l (D) =\sual e_\alpha (D)
\left(l+\frac{D-2}2\right)^{\alpha}.\label{eq:4.3} \eeq
A representation which shows the analytic structure around $s=0$ may
then be given in analogy with Eq.~(\ref{eq:2.7}) (now one defines $\nu
=l+(D-2)/2$)
\beq
\zeta (s) =\sulnu d_l(D) Z_D^{\nu}(s)
+\sum_{i=-1}^{D-1}A_i^D(s).\label{eq:4.4}
\eeq
Here we have used the following definitions. First
\beq
Z_D^{\nu}(s) &=&\sip \iinma\left[\ln I_{\nu}(\nu z) -\nu
\eta\right.\nn\\
& &\left.\hspace{2cm}+\ln\left(\sqrt{2\pi\nu}(1+z^2)^{1/4}\right)-\sum_
{n=1}^{D-1}\frac{D_n (t)}{\nu^n}\right].\label{eq:4.5}
\eeq
The polynomials $D_n(t)$ arise from the asymptotic expansion of $I_{\nu}
(\nu z)$ \cite{abramowitzstegun72}. In detail one has
\beq
I_ {\nu} (\nu z) \sim \frac 1 {\sqrt{2\pi
\nu}}\frac{e^{\nu\eta}}{(1+z^2)^{1/4}}\left[1+\sum_{k=1}^{\infty}\frac{u_
 k (t)}{\nu ^k}\right],\label{eq:4.6}
\eeq
with the recursion relation
\beq
u_{k+1} (t) =\frac 1 2 t^2 (1-t^2)u_k'(t) +\frac 1 8 \int\limits_0^t
d\tau\,\,(1-5\tau^2 )u_k (\tau)\label{eq:4.7}
\eeq
for the polynomials $u_k (t)$. The coefficient functions $D_n (t)$ are
then defined through the cumulant expansion
\beq
\ln\left[1+\sum_{k=1}^{\infty}\frac{u_k (t)}{\nu^k}\right]\sim
\sum_{n=1}^{\infty}\frac{D_n(t)}{\nu^n}\label{eq:4.8}
\eeq
and are easily found with the help of a simple computer program.

The $A_i^D(s)$ have already been determined in \cite{bek} and we give
only their final form for completeness. Introducing
\beq
D_n (t) =\sum_{a=0}^ix_{i,a}t^{i+2a},\label{eq:4.9}
\eeq
they read
\beq
A_{-1}^D (s) &=&\frac{\rzs}{4\sqrt{\pi}\Gamma (s)}\sujnu \lll
(mR)^{2j}\frac{\g j+s-\frac 1 2\right)}{s+j}\label{eq:4.10}\\
& &\times \left[\sual \ead\zeta_H (2j+2s -1-\alpha;(D-2)/2)\right],\nn\\
A_0^D (s) &=&-\frac{\rzs}{4\Gamma (s)}\sujnu \lll (mR)^{2j}\Gamma
(s+j)\label{eq:4.11}\\
& &\times \left[\sual \ead\zeta_H (2j+2s-\alpha;(D-2)/2)\right],\nn\\
A_i^D (s) &=& -\frac{\rzs}{2\Gamma (s)}\sujnu \lll
(mR)^{2j}\label{eq:4.12}\\
& &\hspace{1cm} \times \left[\sual \ead\zeta_H
(-\alpha+i+2j+2s;(D-2)/2)\right] \nn\\
& &\hspace{2cm} \times \suani x_{i,a} \frac{(i+2a)\g s+a+j+\frac i
2\right)}{\g 1+a+\frac i 2\right)}.\nn
\eeq
Their contributions to the functional determinant are easily determined
and are listed in Appendix A, for the dimensions $D=4,5,6$.

For the calculation of $Z_D^{\nu}(s)$ we go on as for the three
dimensional case. We have found
\beq
{Z_D^{\nu}}'(0)&=&\ln\Gamma (\nu +1)+\nu-\nu\ln\nu -\frac 1 2 \ln
(2\pi\nu) +\sum_{n=1}^{D-1}\frac{D_n(1)}{\nu^n}\nn\\
&=&\int\limits_0^{\infty}\left(\frac 1 2 -\frac 1 t +\frac 1 {e^t-1}
\right)\frac{e^{-t\nu}} t+\sum_{n=1}^{D-1}\frac{D_n(1)}{\nu^n}\nn\\
&=&\int\limits_0^{\infty}\left(\sum_{n=1}^{D-1}\frac{D_n(1)}{(n-1)!}t^n
+\frac 1 2 -\frac 1 t +\frac 1 {e^t-1}
\right)\frac{e^{-t\nu}} t.\label{eq:4.13}
\eeq
This is the suitable starting point for the summation of the angular
momentum. Introducing the regularized version of $Z_D(s)$ as in
Eq.~(\ref{eq:2.13}) one obtains
\beq
Z_D'(0,z)&=&\sum_{\alpha =1}^{D-2}\ead
\int\limits_0^{\infty}dt\,\,t^z\frac{e^{-t\frac{D-2}
2}}{1-e^{-t}}\frac{d^{\alpha}}{dt^{\alpha}}\left(\frac 1 t \frac 1
{e^t-1}\right)\nn\\
& &+\sual \ead \left\{\sum_{n\geq \alpha +1}^{D-1}\frac{D_n(1)}{\Gamma
(n-\alpha)}\,\zeta_H\left(z+n-\alpha;\frac{D-2}2\right)\Gamma(z+n-\alpha)
\right.\nn\\
& &\hspace{2cm}+\frac 1 2 (-1)^{\alpha}\alpha ! \zeta_H
\left(z-\alpha;\frac{D-2}2\right)\Gamma (z-\alpha) \label{eq:4.14}\\
& &\left.\hspace{2cm}- (-1)^{\alpha}(\alpha +1) ! \zeta_H
\left(z-\alpha -1;\frac{D-2}2\right)\Gamma (z-\alpha-1)\right\},
\nn
\eeq
which we need at $z=0$. The remaining task is the calculation of the
integral term in Eq.~(\ref{eq:4.14}). It may be given by repeated
differentiation of Eq.~(\ref{eq:2.14}). Defining
\beq
f(1,\beta,\gamma,\nu) =
 \int_0^\infty \frac{x^{\nu -1}  e^{-\gamma
x}}{1-e^{-\beta x}} \, dx  \ = \ \frac{\Gamma (\nu)}{\beta^\nu} \,
\zeta (\nu, \frac{\gamma}{\beta})\label{eq:4.15}
\eeq
and
\beq
f(k,\beta,\gamma,\nu) =
 \int_0^\infty \frac{x^{\nu +k-2}  e^{-[\gamma
+(k-1)\beta]x}}{(1-e^{-\beta x})^k} \, dx, \ \ \ \ k=1,2,
\ldots\label{eq:4.16}
\eeq
we get the recurrence
\beq
f(k, \beta, \gamma , \nu ) &=& \frac 1 {k-1}\frac{\partial}{\partial
\beta} f (k - 1, \beta, \gamma, \nu)
\nn\\& & -\frac{k - 2}{k - 1}
 f(k - 1, \beta, \gamma, \nu  + 1). \label{eq:4.17}
\eeq
The first few cases, $k=2,3,4$, and a general formula for $f(k,\beta
,\gamma,\nu)$ is given in Appendix B. All terms appearing in
Eq.~(\ref{eq:4.14}) and resulting from the integration may be given in
terms of this functions $f(k,\beta ,\gamma ,\nu)$.
In detail this is seen as follows. First of all one may show, that
\beq
\frac{d^{\alpha}}{dt^{\alpha}}\left(\frac 1 t \frac 1 {e^t-1}\right) =
\sum_{k=0}^{\alpha}B_k^{(\alpha)}\sum_{i=1}^{k+1}A_i^{(k)}
\frac{t^{k-1-\alpha}}{(e^t-1)^i},\nn
\eeq
with the recursion relations
\beq
B_0^{(\alpha )}&=& (-1)^{\alpha}\alpha !,\nn\\
B_ {\alpha}^{(\alpha )}&=&1,\nn\\
B_k^{(\alpha )}&=&(k-\alpha )B_k^{(\alpha -1)}+B_{k-1}^{(\alpha -1)},
\qquad \mbox{for}\,\, k=1,...,\alpha -1,\nn
\eeq
for $B_k^{(\alpha )}$, and similar ones for $A _{i}^{(k)}$ \cite{dowrob},
\beq
A_1^{(k)}&=&(-1)^k,\nn\\
A_{k+1}^{(k)}&=&(-1)^k k!,\nn\\
A_i^{(k)}&=&-iA_i^{(k-1)}-(i-1)A_{i-1}^{(k-1)},\qquad \mbox{for}\,\,
2\leq i \leq k,\nn
\eeq
coming from
\beq
\frac{d^k}{dt^k}\frac 1 {e^t-1} =\sum_{i=1}^{k+1}\frac{A_i^{(k)}}
{(e^t-1)^i}.\nn
\eeq
As a result,
\beq
\lefteqn{
\sual \ead \int\limits_0^{\infty}dt\,\, t^z\frac{e^{-t\frac{D-2}2}}
{1-e^{-t}}\frac{d^{\alpha}}{dt^{\alpha}}\left(
\frac 1 t \frac 1 {e^t-1}\right)}\nn\\
&=&\sual \ead\sum_{k=0}^{\alpha}B_k^{(\alpha )}\sum_{i=1}^{k+1} A_i^{(k)}
f\left(i+1,1,\frac{D-2}2,z+k-\alpha -i\right).\nn
\eeq
Thus we have derived all necessary equations, showing that
$Z_D'(0,z)$ may be given solely in terms of $\Gamma$-functions and
Riemann zeta functions and the limit $z\to 0$ may be taken. The results
are listed in Appendix C for $D=4,5,6.$

Adding up the contributions from the asymptotic terms and $Z_D'(0)$,
we have
found the following final results for the zeta function determinants of
the Laplace operator with Dirichlet boundary conditions in $D=4,5$ and
$6$ dimensions,
\beq
\zeta_4'(0) &=& \frac{173}{30240} +\frac 1 {90} \ln 2
- \frac 1 {90} \ln R + \frac 1 3 \zeta_R'(-3)\nn\\
& &-\frac 1 2 \zeta_R'(-2) + \frac 1 6 \zeta_R'(-1),\nn\\
\zeta_5'(0) &=&\frac{47}{9216} + \frac{17}{2880}\ln 2 +
  \frac{17}{5760}\ln R -
\frac 5 {64}\,\zeta_R'(-4)\nn\\
& &+\frac 7 {48}\,\zeta_R'(-3) -
\frac 1 {32}\,\zeta_R'(-2) - \frac 1 {48}\,\zeta_R'(-1),\nn\\
\zeta_6'(0) &=& -\frac{4027}{6486480} -\frac 1 {756}\ln 2
+ \frac 1 {756}\ln R +
\frac 1 {60}\,\zeta_R'(-5) -\frac 1 {24}\,\zeta_R'(-4)\nn\\
& & +\frac 1 {24}\,\zeta_R'(-2) -
\frac 1 {60}\,\zeta_R'(-1).\nn
\eeq
The result for $D=4$ agrees with Dowker and Apps \cite{fundow}.
As is clear from our presentation, {\it any} higher dimension $D$ can be
treated in exactly the same way
without additional problems (they just get a bit
arithmetically cumbersome).

Let us now describe briefly the calculation for Robin boundary
condition. There the starting point is
\beq
\zeta (s,u)=\sulnu d_l(D) \ikma \ln\left[\frac u R J_{\nu} (kR) +
  kJ_{\nu} '(kR)\right].\label{eq:4.21}
\eeq
The relevant asymptotic expansions are (\ref{eq:4.6}) together with
\cite{abramowitzstegun72}
\beq
I_{\nu}' (\nu z) \sim \frac 1 {\sqrt{2\pi
\nu}}\frac{e^{\nu\eta} (1+z^2)^{1/4}} z
\left[1+\sum_{k=1}^{\infty}\frac{v_
 k (t)}{\nu ^k}\right].\label{eq:4.22}
\eeq
The relevant polynomials this time are determined by
\beq
\ln\left[1+\sum_{k=1}^{\infty}\frac{v_k (t)}{\nu^k} +\frac u {\nu} t
\left( 1+\sum_{k=1}^{\infty}\frac{u_k (t)}{\nu^k}\right)\right]\sim
\sum_{n=1}^{\infty}\frac{M_n (t)}{\nu ^n},\label{4.23}
\eeq
where, in analogy to Eq.~(\ref{eq:4.9}), we define
\beq
M_n (t,u) =\sum_{a=0}^{2i} z_{i,a}t^{i+a}.\label{eq:4.24}
\eeq
Continuing as for the $3$-dimensional case, and in analogy to
Eqs.~(\ref{eq:4.4}), (\ref{eq:4.5}), we define
\beq
Z_R^{\nu}(s) &=& \sip \iinma\left\{\ln\left[\frac u R I_{\nu} (z\nu
)+\frac{z\nu}{R}I_{\nu}'(\nu z)\right]\right.\label{eq:4.25}\\
& &\hspace{3cm}\left.
-\ln\left[\frac{\nu}{R\sqrt{2\pi\nu}}e^{\nu\eta}
(1+z^2)^{\frac
1 4}\right]-\sum_{n=1}^{D-1} \frac{M_n(t,u)}{\nu ^n}\right\}.\nn
\eeq
In the limit $m\to 0$, we obtain
\beq
{Z^{\nu}_R}'(s) &=&\int\limits_0^{\infty}dt\,\, \left(
\frac 1
2 -\frac 1 t +\frac 1 {e^t-1}\right)\frac{e^{-t\nu}} t
\label{eq:4.26}\\
& &+\ln\frac{\nu}{u+\nu}+\sum_{n=1}^{D-1}\frac{M_n(t,u)}{\nu^n}\nn
\eeq
and, realizing that
\beq
M_n (1,0)=D_n (1),\qquad M_n (1,u)-M_n (1,0) =(-1)^{n+1}\frac{u^n}
n,\label{eq:4.27}
\eeq
we find, after performing the sum over the angular momentum,
\beq
Z_R'(0) &=& Z_D'(0)\label{eq:4.28}\\
& &+\sulnu \sual \ead \nu^{\alpha} \left\{ -\ln\left( 1+\frac u
{\nu}\right) +\sum_ {n=1}^{D-1} (-1)^{n+1}\frac 1 n \left(\frac u
{\nu}\right)^n\right\}.\nn
\eeq
The remaining sum may once more be done by differentiating and
integrating back, and we end up with
\beq
Z_R'(0) &=& Z_D'(0)\label{eq:4.29}\\
& &+\sual \ead \left\{(-1)^{\alpha +1} \Psi\left(\frac{D-2} 2\right)
\frac{u^{\alpha +1}}{\alpha +1}\right.\nn\\
& &\phantom{\sual \ead \left\{ \right. }
+\sum_{n=\alpha +2}^{D-1} (-1)^{n+1} \zeta_H (n-\alpha;(D-2)/2)
\frac{u^n} n \nn\\
& &\phantom{\sual \ead \left\{ \right. }
-(-1)^{\alpha}\alpha \int\limits_0^u dx\,\, x^{\alpha
-1}\ln\Gamma\left(\frac{D-2}2 +x\right)\nn\\
& &\phantom{\sual \ead \left\{ \right. }
\left.+(-1)^{\alpha}u^{\alpha
}\ln\Gamma\left(\frac{D-2}2 +u\right)\right\}.\nn
\eeq
The results for the special dimensions $D=4,5$ and $6$ are not listed,
because they are essentially given in Eq.~(\ref{eq:4.29}) together with
Appendix C. Let us only note that
\beq
e_1(4) &=& 0,\qquad e_2(4)=1,\nn\\
e_1(5) &=& -\frac 1 {12},\qquad e_2(5) =0,\qquad e_3(5) =\frac 1 3,\nn\\
e_1(6) &=& 0, \qquad e_2(6) =-\frac 1 {12},\qquad
e_3(6) = 0, \qquad e_4(6) =\frac 1 {12}.\nn
\eeq
The contributions of the asymptotic terms are calculated from
\beq
A_{-1}^R (s) =A_{-1}^D(s),\qquad
A_0^R (s) =- A_0^D(s),\label{eq:4.30}
\eeq
and
\beq
A_i^R (s) &=& -\frac{\rzs}{2\Gamma (s)}\sujnu \lll (mR)^{2j}\nn\\
& &\hspace{1cm}  \sual \ead \zeta_H (-1+i+2j+2s;(D-2)/2) \nn\\
& &\hspace{3cm} \times \sum_{a=0}^{2i} z_{i,a} \frac{(i+a)\g
s+j+\frac{a+ i} 2\right)}{\g 1+\frac {a+i} 2\right)}.\label{eq:4.31}
\eeq
They are collected in Appendix A.

Summing up, we have found the following final results for the functional
determinant with Robin boundary conditions:
\beq
\zeta_4'(0,u)&=&
{{11}\over {4320}} + {u\over {30}} - {{5\,{u^2}}\over {12}} -
  {{{u^3}}\over 3} + {{\ln (2)}\over {90}} + {{{u^3}\,\ln(2)}\over 3}
\nn\\
& &-  {{\ln (R)}\over {90}} - {{{u^3}\,\ln (R)}\over 3} +
  {{\zeta_R'(-3)}\over 3} + {{\zeta_R'(-2)}\over 2} +
  {{\zeta_R'(-1)}\over 6}\nn\\
& &+u^2\ln\Gamma (1+u) -2\int\limits_0^u dx\,\,x\ln\Gamma (1+x),\nn\\
\zeta_5'(0,u)&=&
-{{61}\over {46080}} - {{11\,u}\over {576}} - {{{u^2}}\over {16}} +
  {{11\,{u^3}}\over {72}} + {{{u^4}}\over {24}} \nn\\
& & + {{7\,\ln (2)}\over {720}} - {{17\,\ln (R)}\over {5760}} -
  {{{u^2}\,\ln (R)}\over {24}} + {{{u^4}\,\ln (R)}\over {12}} \nn\\
& &-  {{5\,\zeta_R'(-4)}\over {64}} - {{7\,\zeta_R'(-3)}\over {48}} -
  {{\zeta_R'(-2)}\over {32}} + {{ \zeta_R'(-1)}\over {48}}\nn\\
& &-\frac 1 {12} \int\limits_0^udx\,\,\ln\Gamma\left(\frac 3 2 +x\right)
+\int\limits_0^udx\,\,x^2\ln\Gamma\left(\frac 3 2 +x\right)\nn\\
& &+\frac 1 {12} u \ln\Gamma\left(\frac 3 2 +u\right)
-\frac 1 3 u^3\ln\Gamma\left(\frac 3 2 +u\right),\nn\\
\zeta_6'(0,u)&=&
-{{9479}\over {32432400}} - {u\over {315}} + {{517\,{u^2}}\over {15120}} +
  {{83\,{u^3}}\over {1512}} - {{19\,{u^4}}\over {480}}\nn\\
& &-  {{{u^5}}\over {45}} - {{\log (2)}\over {756}} -
  {{{u^3}\,\ln (2)}\over {36}} + {{{u^5}\,\ln (2)}\over {60}} +
  {{\ln (R)}\over {756}} + {{{u^3}\,\ln (R)}\over {36}}\nn\\
& &-  {{{u^5}\,\ln (R)}\over {60}} + {{\zeta_R'(-5)}\over {60}} +
  {{\zeta_R'(-4)}\over {24}} - {{\zeta_R'(-2)}\over {24}} -
  {{\zeta_R'(-1)}\over {60}}\nn\\
& &+\frac 1 6 \int\limits_0^u dx\,\, x\ln\Gamma (2+x) -\frac 1 3
\int\limits_0^u dx\,\, x^3\ln\Gamma (2+x)\nn\\
& &-\frac 1 {12} u^2\ln\Gamma (2+u) +\frac 1 {12}
u^4\ln\Gamma (2+u).\nn
\eeq
The detailed dependence of $\zeta '(0,u)$ on the parameter $u$ for dimensions
$D=4,5,6$ is given in Fig. 1, for $R=1$.

Finally, we are left with the task of the calculation of the zeta
function determinant for Neumann boundary conditions.
As we have already seen, the $l=0$ term corresponding to $\nu =(D-2)/2$
has to be treated separately. Employing the same procedure as for $3$
dimensions, one finds
\beq
\frac d {ds} \left( \zeta^{l=0} (s,1-D/2) -\zeta^{l=0}
(s,u)\right)\left.\right|_{s=0} &=&\nn\\
& &\hspace{-4cm} \sual \ead
\left(\frac{D-2} 2 \right)^{\alpha} \left[-2\ln R +\ln (D/2) +\ln (
(D-2)/2 +u)\right],\nn
\eeq
which results in the following final results for Neumann boundary conditions,
\beq
{\zeta_4}' (0,-1) &=&
 -{{493}\over {4320}} + {{61\,\ln 2}\over {90}} -
   {{151\,\ln R}\over {90}}\nn\\& & + {{\zeta_R'(-3)}\over 3} +
   {{\zeta_R'(-2)}\over 2} + {{\zeta_R'(-1)}\over 6}\nn\\
& &+2\int\limits_0^1dx\,\,(x-1)\ln\Gamma (x),\nn
\eeq
\beq
{\zeta_5}'(0,-3/2) &=&
 -{{19261}\over {46080}} - {{713\,\ln 2}\over {720}} + \ln 5 -
   {{9647\,\ln R}\over {5760}}\nn \\
& & - {{5\,{\zeta_R}'(-4)}\over {64}} -
   {{7\,{\zeta_R}'(-3)}\over {48}} - {{{\zeta_R}'(-2)}\over {32}} +
   {{{\zeta_R}'(-1)}\over {48}}\nn\\
& &+\frac 1 {12} \int\limits_0^{3/2}dx\,\,\ln\Gamma (x) -
\int\limits_0^{3/2} dx\,\,\left(x-\frac 3 2 \right)^2 \ln\Gamma (x), \nn
\eeq
\beq
{\zeta_6}' (0,-2) &=&
 -{{7087979}\over {32432400}} - {{1181\,\ln 2}\over {3780}} +
   \ln 3 - {{6379\,\ln R}\over {3780}} \nn\\ & &+
   {{{\zeta_R}'(-5)}\over {60}} + {{{\zeta_R}'(-4)}\over {24}} -
   {{{\zeta_R}'(-2)}\over {24}} - {{{\zeta_R}'(-1)}\over {60}}\nn\\
& &-\frac 1 6 \int\limits_0^2dx\,\,(x-2)\ln\Gamma (x) +\frac 1 3
\int\limits_0^2 dx\,\,(x-2)^3\ln\Gamma (x). \nn
\eeq
This terminates the explicit calculation for all kind of
 boundary conditions and for dimensions $D=3,4,5,6$ (that can be
extended immediately to any
dimension $D>6$). Once more the results may be given in an
alternative form by using Eq.~(\ref{lngam}), but this makes little
difference for numerical calculations.

Finally, we are left with the case $D=2$.
\section{Zeta function determinants on the $2$-dimensional ball}
\setcounter{equation}{0}
For the $2$-dimensional ball the procedure has to be changed slightly.
Here the degeneracy of every $l\geq 1$ is $2$, $l=0$ has to be counted
only once. Due to the presence of this term $l=0$, the starting point
Eq.~(\ref{eq:2.7})-(\ref{eq:2.7d}) is not valid any more and may be applied
only to $l\geq 1$. The $l=0$ term may be treated as before for the
Neumann boundary conditions. We shall not give any further details for
this case but only write down the final results
---which on the other hand have appeared in part in
the literature \cite{weisberger87}. We quote them
for completeness.

For Dirichlet boundary conditions we have
\beq
\zeta_D' (0) =\frac 5 {12} +2\zeta_R' (-1) +\frac 1 2 \ln \pi +\frac 1 6
\ln 2 +\frac 1 3 \ln R.\nn
\eeq
The zeta function determinant for general Robin boundary conditions
reads
\beq
\zeta_R'(0) =-\frac 7 {12} +\frac 1 3 \ln  R +2\zeta_R '(-1) -\frac 5 6
\ln 2 -\frac 1 2 \ln \pi +2u\ln(2/R) -\ln u +2\ln\Gamma (1+u) .\nn
\eeq
Finally, the results for Neumann boundary condition is
\beq
\zeta_N'(0) =-\frac 7 {12} -\frac 5 3 \ln R+2\zeta_R'(-1) +\frac 1 6 \ln
2 -\frac 1 2 \ln \pi .\nn
\eeq
And this concludes the list of examples of zeta function determinants on
the ball that we had promised to consider.
\section{Conclusions}
In this paper we have developed a systematic approach for the
calculation of
functional determinants of (elliptic differential) operators, which is
very useful in all cases when
the basis of functions ---constrained by the equations cooresponding to
the boundary conditions--- is known.
Using our approach we have calculated the zeta function determinant of
the
Laplacian on the ball for Dirichlet, Neumann and the general Robin boundary
conditions. Explicit results for dimensions $D\leq 6$ have been given.
All
necessary formulas to iterate the calculation of $\zeta ' (0)$  and to
obtain it in any dimension, by means of a simple computer program, have
been given explicitly.

An extension of the present work to higher-dimensional bundles, such as
those for spinors and vectors, is envisaged. This should allow to
establish connections with recent work dealing with mixed boundary
conditions
\cite{vassilevich,mosspoletti94,espositokamenshchikmishakovpollifrone94a}.

\vspace{1cm}

\noindent {\bf Comment:}
Using a different approach, based on work by Moss \cite{moss89} and
Voros
\cite{voros87}, Dowker has also considered the calculation of
heat-kernel coefficients and
functional determinants for the Laplace operator on the ball with Robin
boundary conditions \cite{dowrob}. We are indebted with him for
interesting and fruitful correspondence during the course of the
present work.

\vspace{1cm}

\noindent {\large \bf Acknowledgments}

KK would like to thank the members of the Department ECM, Barcelona
University, for their kind hospitality.
This investigation has been partly supported by
the Alexander von Humboldt foundation (Germany), by
DGICYT (Spain), project
PB93-0035, by the german-spanish program Acciones Integradas,
project HA94-072 and by DFG under contract number Bo 1112/4-1.

\vspace{1cm}

\begin{appendix}
\section{Appendix: Contributions to the zeta function determinant from the
asymptotic terms}
\setcounter{equation}{0}
In this appendix we list the contributions of the asymptotic terms to
the zeta function determinant for Dirichlet and Robin boundary
conditions.
\subsection{Dirichlet boundary condition}
For Dirichlet boundary conditions we find
\beq
(Asym)_D(D=4)&=&-\frac{83}{6048} - \frac{\gamma}{360} +\frac{\ln 2}{90}
- \frac{\ln R}{90} -
\zeta_R'(-3) -\frac{\zeta_R'(-2)}{2},\nn\\
(Asym)_D(D=5)&=&
\frac{47}{9216}+\frac{\pi^2}{8640} -\frac{11}{5760}\ln 2 +
\frac{17}{5760}\ln R \nn\\
& &+\frac 5 {16} \zeta_R' (-4)+\frac 7 {48} \zeta_R ' (-3) -
\frac 1 {16} \zeta_R ' (-2) -\frac 1 {48} \zeta_R ' (-1),\nn\\
(Asym)_D(D=6)&=&\frac{40091}{25945920} +\frac{\gamma}{3360} - \frac{\ln
2}{756} + \frac{\ln R}{756} -
\frac{\zeta_R(3)}{15120} -\frac{\zeta_R'(-5)}{12}\nn\\
& &-\frac{\zeta_R'(-4)}{24} +
\frac{\zeta_R'(-3)}{12} +\frac{\zeta_R'(-2)}{24}.\nn
\eeq
\subsection{Robin boundary condition}
For Robin boundary conditions the final results read
\beq
(Asym)_R(D=4) &=& -\frac{73}{4320} - \frac 1 {360}\gamma +\frac 1 {90}
\ln 2
-\frac 1 {90}\ln R -\zeta_R'(-3) +\frac 1 2\zeta_R'(-2)\nn\\
& & +\frac u {30} - \frac 5 {12} u^2 - \frac 1 3 u^3 -
\frac 1 3 \gamma u^3 +\frac 1 3 u^3\ln 2 -
\frac 1 3 u^3\ln R,\nn\\
(Asym)_R(D=5) &=& -\frac{61}{46080} + \frac{\pi^2}{8640} -
+ \frac{11}{5760} \ln 2-\frac{17}{5760}\ln R\nn\\
& &+\frac 5 {16}\,\zeta_R'(-4) -
\frac 7 {48}\,\zeta_R'(-3) -\frac 1 {16}\,\zeta_R'(-2) +
\frac 1 {48}\,\zeta_R'(-1)\nn\\
& &-\frac{11}{576}u + \frac 1 {48} u^2 - \frac 1 {24}\gamma u^2 +
 \frac 1 {24}u^3 + \frac 1 {72} \pi^2 u^3 + \frac 1 {24}u^4\nn\\
& &+\frac 1 {12} \gamma u^4 -
\frac 1 {12}u^2\ln 2 +\frac 1 6 u^4\ln 2 -
\frac 1 {24}u^2\ln R +
\frac 1 {12} u^4\ln R - \frac 7 {48}u^4\zeta_R(3),\nn\\
(Asym)_R(D=6) &=&\frac{243079}{129729600} +\frac 1 {3360}\gamma
-\frac 1 {756} \ln 2 -\frac 1 {15120} \zeta_R(3)\nn\\
& &-\frac 1 {12} \zeta_R'(-5) +
\frac 1 {24}\,\zeta_R'(-4) +\frac 1 {12}\,\zeta_R'(-3) -
\frac 1 {24}\,\zeta_R'(-2)\nn\\
& &-\frac 1 {315}  u + \frac{517}{15120}u^2 +
\frac{41}{1512}u^3 + \frac 1 {36} \gamma u^3 - \frac 3 {160} u^4 \nn\\
& &-\frac 1 {288} \pi^2 u^4 -
\frac 1 {45} u^5 - \frac 1 {60} \gamma u^5 -
\frac 1 {36} u^3 \ln 2 +
\frac 1 {60} u^5 \ln 2 + \frac 1 {756}\ln R \nn\\& &+ \frac 1
{36} u^3\ln R -\frac 1 {60}u^5\ln R
+ \frac 1 {60} u^5\zeta_R(3).\nn
\eeq
\section{Appendix: Detail of the calculation of $Z_D'(0)$}
\setcounter{equation}{0}
Using the recurrence relation (\ref{eq:4.17}), the function
$f(k,\beta ,\gamma , \nu)$ for $k=2,3$ and $4$ read,
\beq
f(2,\beta,\gamma ,\nu ) &=&
 \int_0^\infty \frac{x^\nu  e^{-(\gamma + \beta)x }}{(1-e^{-\beta x})^2}
\, dx \nn \\ &=&
\frac{\Gamma (\nu +1)}{\beta^{\nu +1}} \left[
      {\zeta}(\nu , {\gamma \over \beta}) -\frac{\gamma}{\beta}
 \,{\zeta}(\nu +1 , {\gamma \over \beta})
\right], \\
f(3,\beta,\gamma ,\nu ) &=& \int_0^\infty \frac{x^{\nu +1}  e^{-(\gamma
+2\beta )x }}{(1-e^{-\beta x})^3} \, dx
\nn  \\ &&   \hspace{-25mm} =
\frac{\Gamma (\nu +2)}{2 \beta^{\nu +2}}
       \left[ {\zeta}(\nu , {\gamma \over \beta}) -
         (1+2\frac{\gamma}{\beta})\,{\zeta}(\nu +1 , {\gamma
\over \beta})+ \frac{\gamma}{\beta}
         {(1+\frac{\gamma}{\beta})}\,{\zeta}(\nu +2 , {\gamma
\over \beta})
\right], \\
f(4,\beta,\gamma ,\nu ) &=& \int_0^\infty \frac{x^{\nu +2}  e^{-(\gamma
+3\beta )x }}{(1-e^{-\beta x})^4} \, dx
\nn  \\ &&   \hspace{-25mm} =
\frac{\Gamma (\nu +3)}{2 \beta^{\nu +3}}
       \left[ {\zeta}(\nu , {\gamma \over \beta}) -
        3(1+\frac{\gamma}{\beta})\,{\zeta}(\nu +1 , {\gamma
\over \beta})+
        [2+6\frac{\gamma}{\beta}+3(\frac{\gamma}{\beta})^2]
\,{\zeta}(\nu +2 , {\gamma \over \beta}) \right. \nn \\
&& \left. -\frac{\gamma}{\beta} [2+3\frac{\gamma}{\beta}
+(\frac{\gamma}{\beta})^2] \,{\zeta}(\nu +3 , {\gamma \over \beta})
 \right].
\eeq
For the general formula one has
\beq
f(k,\beta,\gamma ,\nu ) =
\frac{\Gamma (\nu +k -1)}{(k-1)! \, \beta^{\nu +k-1}} \sum_{j=0}^{k-1}
(-1)^j c_{k-1,j} (\frac{\gamma}{\beta}) \, \zeta (\nu+k,
\frac{\gamma}{\beta}),
\eeq
where
\beq
c_{k-1,0} (\frac{\gamma}{\beta})&=&1, \
c_{k-1,1} (\frac{\gamma}{\beta})=
\frac{\gamma}{\beta} + (\frac{\gamma}{\beta} +1) + \cdots +
(\frac{\gamma}{\beta}+k-2), \ \ldots, \nn\\
c_{k-1,k-1} (\frac{\gamma}{\beta})&=& \frac{\Gamma
(\frac{\gamma}{\beta} +k-1)}{\Gamma(\frac{\gamma}{\beta})}.\nn
\eeq

\section{Appendix: Contribution of $Z_D(s)$ to the zeta function determinant}
\setcounter{equation}{0}
In this appendix we list the contributions of $Z_D(s)$ to the zeta
function determinant for dimensions $D=4,5$ and $6$.
They are
\beq
{Z_D^4}'(0) &=&
\frac 7 {360} +\frac 1 {360} \gamma +\frac 4 3 \zeta_R ' (-3) +\frac 1 6
\zeta_R ' (-1),\nn\\
{Z_D^5}'(0) &=& -\frac{\pi^2}{8640} + \frac 1 {128}\ln 2 -
\frac{25}{64}\,\zeta_R'(-4) +
\frac 1 {32}\,\zeta_R'(-2),\nn\\
{Z_D^6}'(0)&=& -\frac{131}{60480} -\frac{\gamma}{3360} +
\frac 1 {15120} \zeta_R(3) +
\frac 1 {10}\,\zeta_R'(-5) -\frac 1 {12}\,\zeta_R'(-3) -
\frac 1 {60}\,\zeta_R'(-1).\nn
\eeq

\end{appendix}

\newpage

\noindent{\large \bf Figure caption:}
\bigskip

\noindent{\bf Figure 1:}

Plot of the dependence of $\zeta' (0,u)$ on the parameter $u$
 for $R=1$ and for dimensions $D=3,4,5,6$. Notice the divergence that
apppears for $u=1-D/2$ in each dimension, corresponding to the case of
Neumann boundary conditions.

\end{document}